\documentclass[epj]{webofc}
\usepackage[varg]{txfonts}   
\usepackage{graphicx,color} 
\usepackage{bm}       
\usepackage{amsmath}  
\usepackage{amssymb}
\woctitle{21st International Conference on Few-Body Problems in Physics}
\begin{document}
\title{Electromagnetic structure of light nuclei}
\author{Saori Pastore\inst{1,2}\fnsep\thanks{\email{saori@lanl.gov}} }

\institute{Department of Physics and Astronomy, University of South Carolina, Columbia, SC 29208, USA 
\and
Theoretical Division, Los Alamos National Laboratory, Los Alamos, NM 87545, USA}

\abstract{The present understanding of nuclear electromagnetic properties including 
electromagnetic moments, form factors and transitions in nuclei with
A $\le$ 10 is reviewed. Emphasis is on calculations based on nuclear Hamiltonians
that include two- and three-nucleon realistic potentials, along with
one- and two-body electromagnetic currents derived from a chiral effective
field theory with pions and nucleons.
}

\maketitle
\section{Introduction}
\label{intro}

A major goal in nuclear physics is to understand nuclear structure and dynamics in terms of 
underlying interactions occurring between individual nucleons. Studies grounded on this basic 
picture of the nucleus are referred to as {\it ab initio}. An exceptionally powerful tool to 
asses the validity of our theoretical models is to investigate nuclear electromagnetic (e.m.) 
observables, such as ground state properties, {\it e.g.}, e.m. moments and form 
factors, as well as e.m. reactions, {\it e.g.}, photo- and electro-induced reactions. 
In these processes, external e.m. probes interact with the nuclear charge and current 
distributions with a strength characterized by the fine-structure constant $\alpha\sim 1/137$.
The small value of the fine-structure constant allows for a perturbative treatment of the 
e.m. interaction, while non-perturbative physics pertain only to the nuclear target. 
For light nuclei, terms that go beyond the leading order contribution in the Z$\alpha$-expansion 
(where $Z$ is the number of protons) can be safely disregarded, leaving
us with relatively simple reaction mechanisms and manageable formal expressions. For example, 
at leading order, the cross section associated with inclusive electro-nucleus scattering 
processes is factorized into the leptonic tensor, 
which is completely specified by the measured electron's kinematic variables, and the hadronic 
one associated with the nuclear target, and proportional to matrix elements squared of 
the nuclear e.m. charge and current operators. A clear connection between measured 
quantities, {\it i.e.}, cross sections, and calculated matrix elements is then realized. 
Experimental data of e.m. observables are, in most cases, known with great accuracy
providing us with viable and strong constraints on our models. Likewise, for light nuclei,
theoretical calculations are affected by relatively small statistical errors because
for these systems the many-body problem can be solved exactly or within controlled approximations.
This allows for solid comparisons between experimental data and theoretical predictions. 

\begin{figure}
\centering
\includegraphics[width=6cm,clip]{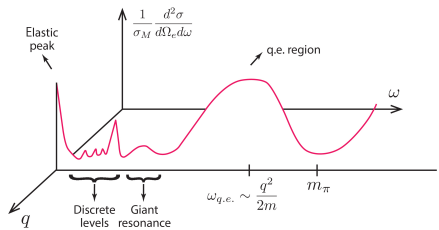}
\caption{Schematic representation of the double differential cross section for electron scattering off nuclei
         at fixed value of momentum transfer.}
\label{fig-1}       
\vspace*{-0.2in}
\end{figure}

In Fig.~\ref{fig-1}, a cartoon picture of the double differential cross section for electron
scattering off nuclei is represented. Different values of energy $\omega$ transferred to the system,
correspond to different excitation energies of the nucleus.
By varying $\omega$, we can access the ground state (elastic peak), low-lying (discrete) nuclear 
excited states, giant resonance modes, and the quasi-elastic energy region up to the pion-production
threshold. For each value of excitation energy $\omega$, one can study 
the matrix elements' behavior as a function of the momentum $|{\bf q|}$  transferred to the nucleus. 
In particular, by varying $|{\bf q|}$ one can explore the e.m. charge and current distributions
with a spatial resolution $\propto 1/|{\bf q}|$. In this talk, I will focus on {\it ab initio} 
calculations of ground state nuclear e.m. properties, that is e.m. moments and elastic form factors, 
as well as widths of e.m. transitions occurring between low-lying nuclear states. These
studies have been recently reported in a topical review on e.m. reactions on light 
nuclei~\cite{review2014}, where more details and references to original articles can be found. 
Recent developments on theoretical {\it ab initio} investigations
on other very interesting e.m. processes in light nuclei, such as photo-absorption and radiative capture reactions, 
Compton scattering, sum rules ..., are well represented in this conference, see, {\it e.g.},
contributions by X.~Zhang, J.~Dohet-Eraly, H.~Griesshammer, M.~Miorelli, S.~Bacca, N.~Barnea, 
D.~Rozpedzik, and A.~Lovato in these proceedings. 

A theoretical understanding and control of nuclear e.m. structure and dynamics is 
a necessary prerequisite for studies on weak induced reactions, such as neutrino-nucleus
interactions.  The experimental data acquisition for this kind of processes is comparatively
more involved owing to the tinier cross sections and to the fact that
neutrinos are chargeless particles and, thus, they are hard to collimate
and detect. An important advance in this direction has recently been carried out by Lovato
and collaborators~\cite{Lovato14}, and for a status report on {\it ab initio} calculations 
of weak response functions in $^4$He and  $^{12}C$ I refer to the plenary talk of 
A.~Lovato (the associated contribution can be found  in these proceedings). Moreover, 
a theoretical understanding of the structure and dynamics of light nuclei is a
necessary prerequisite for research projects aimed at studying larger nuclear systems. 
For these reasons, it is imperative to first validate our theoretical understanding 
of e.m.~reactions on light nuclei.

\section{Nuclear Hamiltonians and electromagnetic currents  }
\label{sec-0}

In the {\it ab initio} framework, the nucleus is described as a system made of $A$ non-relativistic 
point-like nucleons interacting among each other via many-body forces and its energy is approximated by the 
following Hamiltonian:
\begin{equation}
 H = \sum_i K_i + \sum_{i<j} v_{ij} + \sum_{i<j<k} V_{ijk} \ , 
\end{equation}
where  $K_i$ is the non-relativistic single-nucleon kinetic energy, while 
$v_{ij}$ and $V_{ijk}$ are two-nucleon (NN) and three-nucleon (3N) potentials, respectively.
Implicit in the equation above is the assumption the four-nucleon forces and higher order terms 
in the many--body expansion are suppressed. The NN and 3N potentials are phenomenological in nature
in that they involve a number of parameters---subsuming underlying Quantum Chromodynamics (QCD)
effects---that are fixed by fitting experimental data.  For example, NN potentials are
constrained to reproduce a large number of NN scattering data, along with the deuteron binding energy. 
Nuclear forces belonging to this class of highly accurate nuclear potentials are referred to as `realistic'. 
Most realistic potentials describe the long range ($\propto 1/m_\pi$ where $m_\pi$ is the pion mass) part  
of the nuclear interaction in terms of one-pion-exchange interaction mechanisms. Different 
dynamical schemes are implemented to account for intermediate and short range effects, among which 
multiple-pion-exchange, contact interactions, heavy-meson-exchange, or excitations of nucleons into
virtual $\Delta$-isobars. Here, the realistic potentials utilized to solve the Schr\"odinger equation
$H|\Psi\rangle=E|\Psi\rangle$ (where $|\Psi \rangle$ is the nuclear wave function) are the Argonne 
$v18$~\cite{AV18} (AV18) NN potential in combination with either the Urbana IX~\cite{Pudliner95} 
or Illinois-7~\cite{IL7} 3N potentials, as well as combinations of NN and 3N potentials derived 
from chiral effective field theory  ($\chi$EFT)\cite{Machleidt11,Epelbaum12,Gazit09,Navratil07b}.


\begin{figure}
\centering
\includegraphics[height=.20\textheight]{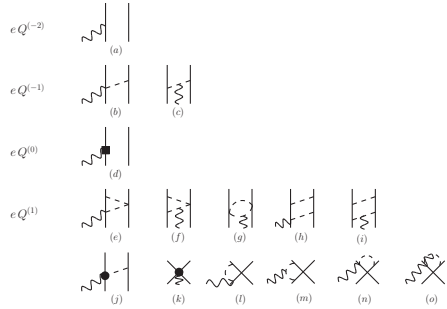}
\caption{Diagrams illustrating one- and two-body EM currents entering at LO ($e\, Q^{-2}$),
NLO ($e\, Q^{-1}$), N2LO  ($e\, Q^{\,0}$), and N3LO ($e\, Q^{\,1}$).  Nucleons, pions,
and photons are denoted by solid, dashed, and wavy lines, respectively.}
\label{fig-2}
\vspace*{-0.2in}
\end{figure}

Nuclear charge ($\rho$) and current ({\bf j}) operators describe the interactions of nuclei 
with external e.m. probes. They are also expanded in a series of many-body operators as
\begin{eqnarray}
\rho    &=& \sum_i {\rho}_i({\bf q}) + \sum_{i<j} {\rho}_{ij}({\bf q}) + \dots\ , \\
\nonumber
{\bf j} &=&  \sum_i {\bf j}_i({\bf q}) + \sum_{i<j} {\bf j}_{ij}({\bf q}) +\dots  \ ,
\end{eqnarray}
where ${\bf q}$ is the momentum transferred to the nucleus. In Impulse Approximation (IA),
that is retaining only one-body operators in the equations above, nuclear e.m. charge and 
current distributions are simply the sums of those associated with individual protons and 
neutrons. The non-relativistic charge operator for point-like nucleons is simply the proton charge, while
the nucleon current consists of a convection term associated with the current generated by 
moving protons and a spin-magnetization term associated with the spins of both protons and neutrons. 
The IA picture of the nucleus is, however, incomplete as it fails to explain, {\it e.g.}, the
measured magnetic moments of light nuclei. Corrections that account for processes in which external
e.m. probes couple to pairs of interacting nucleons, described by two-body current operators, 
need to be incorporated in the theoretical {\it ab initio} description. Meson-exchange 
currents (MEC)---postulated in the '40s by Villars~\cite{Villars47} and 
Miyazawa~\cite{Miyazawa51}---follow naturally once meson-exchange mechanism are invoked 
to describe interactions between individual nucleons. They  account for processes in which 
the external e.m. probe couples with mesons being exchanged between nucleons. The first evidence
of meson-exchange effects in light nuclei can be traced back to the 1972 work by  Riska and 
Brown~\cite{Riska72}, in which MEC were found to provide the missing $10\%$ correction to the IA value 
necessary to reach agreement between the calculated and the measured cross sections for the
radiative capture of proton on neutron at thermal neutron energies. Since then, MEC have evolved
into highly sophisticated and accurate currents. In their most recent formulation~\cite{Marcucci05,Marcucci08},
in order to assure consistency between nuclear forces and e.m. currents, MEC are constructed from realistic 
NN and 3N potentials so as to satisfy the continuity equation. Addition of these MEC corrections to the 
IA picture successfully explains a wide number of e.m. nuclear observables in light nuclei~\cite{Carlson98,Lovato13}.

Recent years have witnessed the tremendous development and success of $\chi$EFT~\cite{Weinberg90,Weinberg91,Weinberg92}
that reinforces and grounds the achievements of conventional theoretical approaches.
The relevant degrees of freedom of nuclear physics are bounds states of QCD, {\it i.e.}, 
pions, nucleons, and $\Delta$'s, $\dots$. On this basis, their dynamics is completely determined
by that associated with the underlying degrees of freedom of quarks and gluons, that is QCD. 
However, at low energies, QCD does not have a simple solution because the strong coupling 
constant becomes too large and perturbative techniques cannot be applied to solve it.
$\chi$EFT is a low-energy approximation of QCD  valid in the energy regime where 
the typical momenta involved, generically indicated by $Q$, 
are such that $Q\ll\Lambda_\chi \sim 1$ GeV, where $\Lambda_\chi$ is the chiral-symmetry breaking scale.
$\chi$EFT provides us with effective Lagrangians describing the interactions between pions, nucleons,
and $\Delta$'s that preserve all the symmetries, in particular chiral symmetry,
exhibited by the underlying theory of QCD at low-energy. These effective interactions, and the transition amplitudes derived from them, 
can be expanded in powers of the small expansion parameter $Q/\Lambda_\chi$, restoring, in practice, 
the possibility of applying perturbative techniques also in the
low-energy regime. The unknown coefficients of this expansion in small 
momenta---referred to as low-energy constants (LECs)---while being tied to QCD effects
and therefore attainable from QCD calculations, are, in practice, fixed 
by comparison with the experimental data. Due to the chiral expansion,
it is then possible to evaluate nuclear observables to any degree $\nu$ of desired accuracy, 
with an associated theoretical error roughly given by $(Q/\Lambda_\chi)^{(\nu+1)}$. 
This calculational scheme has been widely utilized to study both nuclear forces and nuclear
electroweak currents. The many-body operators emerging from direct evaluations
of the transitions amplitudes with interactions provided by $\chi$EFT Lagrangians  
involve multiple-pion exchange operators, as well as contact-like interaction terms. 
Nuclear two-- and three--body interactions were first investigated in the late `90s
by Ord\`o\~nez, Ray, and van Kolck using a $\chi$EFT with pions and nucleons~\cite{vanKolck94,Ordonez92,Ordonez96}.
Currently, chiral NN (3N) potentials commonly used in {\it ab initio} calculations include
up to next-to-next-to-next-to leading order or N3LO  (next-to-next-to leading order or N2LO) 
corrections in the chiral expansion~\cite{Epelbaum09,Epelbaum12,Machleidt11}.

Vector e.m. currents have been first derived  from a $\chi$EFT with pions and nucleons by Park, Min,
and Rho in Ref.~\cite{Park96}. The resulting operators account for two-pion exchange terms entering
at N3LO in the chiral expansion. These currents have been utilized in a number of so called `hybrid' 
calculations\footnote{In this kind of calculations, matrix elements of the chiral e.m. charge and current operators
are evaluated in between  wave functions obtained from conventional realistic potentials, as opposed 
to potentials derived  consistently from $\chi$EFT.}  
of nuclear e.m. observables, including magnetic moments and M1 properties of $A=2$--$3$ nuclei
and radiative capture cross sections in $A=2$--$4$ systems~\cite{Song07,Song09,Lazauskas09}. 
More recently, $\chi$EFT e.m.~currents and charge have been derived up to one loop contributions included
within two different implementations of time ordered perturbation theory: one is by the JLab-Pisa 
group (see Refs.~\cite{Pastore08,Pastore09,Pastore11,Piarulli12}) and the other one is by the 
Bochum-Bonn group (see Refs.~\cite{Kolling09,Kolling11}). In this talk, I focus on results
obtained using chiral e.m. currents, and compare, where possible, different theoretical evaluations
against the experimental data. For results based on conventional e.m. MEC currents I refer the
reader to the review articles of Refs.~\cite{Carlson98,review2014} and references therein.

Before I proceed presenting applications to e.m. observables, I will briefly describe
the e.m. operators as they emerge from a $\chi$EFT with pion end nucleons. I will start off 
with the vector e.m. current which is diagrammatically represented in Fig.~\ref{fig-2}.
The leading order (LO) contribution to the e.m. current illustrated in panel (a) is simply given by the non-relativistic one-body 
current used in IA calculations, while the N2LO one-body operator of panel (d) is a relativistic
correction to the LO IA current. Currents of one- and two-pion range, describing long
and intermediate range dynamics, enter at NLO and N3LO---panels (b), (c), (e)---(j).
Short-range dynamics is encoded by the contact currents of panel (k). Unknown LECs
enter the tree-level diagram of panel (e) and contact currents of panel (k). LECs
entering the contact terms are of two kinds, namely minimal and non-minimal. 
The former enter also the NN chiral potential at NLO, and can then be constrained
to NN scattering data; the latter need to be fixed from e.m. experimental data.
A common procedure implemented to reduce the number of unknown non-minimal LECs
(there are 5 of them) is to impose that the two LECs entering the 
isovector part of the tree-level current illustrated in panel (e) are in fact
saturated by the $\Delta$-couplings entering the $\Delta$ transition e.m. 
current~\cite{Pastore09,Piarulli12}. The remaining three LECs are commonly fixed
so as to reproduce the magnetic moments of the deuteron, triton, and $^3$He~\cite{Piarulli12}.

Early investigations on the e.m. charge operator in $\chi$EFT have been carried out in
Refs.~\cite{Walzl01,Phillips03,Phillips07}, and more recently loop corrections have been
derived in Ref.~\cite{Kolling09}, and subsequently Ref.~\cite{Pastore11}. In closing, we 
note that the structure of the charge operator is quite different from that of the vector e.m. current.
Two-body corrections, in this case, are expected to be relatively small.
In fact, leading two-body operators of one-pion range are suppressed as 
they enter at N3LO (as opposed to NLO as seen in the case of the vector
currents), while there are no free LECs entering the charge operator~\cite{Pastore11}. 

\section{Deuteron, $^3$He and $^3$H electromagnetic form factors}
\label{sec-1}

\begin{figure}
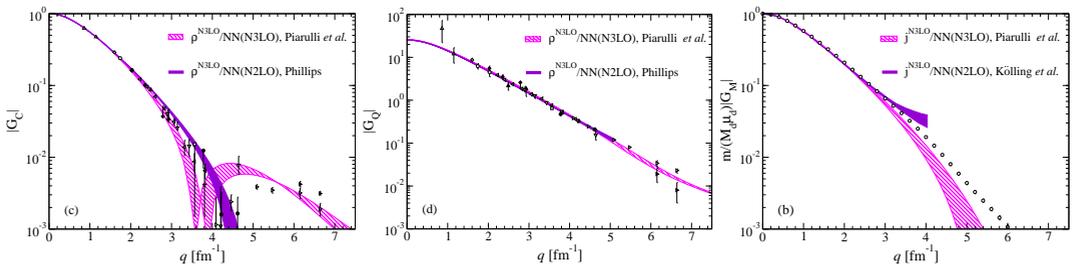

\begin{minipage}{0.23\textheight}
\includegraphics[width=\linewidth]{pastores_fig3.eps}
\end{minipage}
\begin{minipage}{0.23\textheight}
 \includegraphics[width=\linewidth]{pastores_fig4.eps}
\end{minipage}
\begin{minipage}{0.23\textheight}
\vspace*{0.in}
\includegraphics[width=\linewidth]{pastores_fig5.eps}

\end{minipage}
\caption{  {\bf Left}: Deuteron charge form factor from Ref.~\cite{Piarulli12} (magenta hatched band)
            and Ref.~\cite{Phillips07} (purple solid bands) compared with experimental data (black points).
           {\bf Middle}: Same as left panel but for deuteron quadrupole form factor.
           {\bf Right}: Deuteron magnetic form factor from Ref.~\cite{Piarulli12} (magenta hatched band) and 
           Ref.~\cite{Kolling12} (purple solid band)  compared with experimental data (open circles).
           }
\label{fig-3}
\vspace*{-0.4in}
\end{figure}

For $A= 2$-$3$ nuclei, theoretical calculations performed by different groups are available,
which makes it possible to compare them not only with the experimental data but also between 
themselves to test the solidity of the {\it ab initio} prescription. The left and middle
panels of Fig.~\ref{fig-3}, show the deuteron charge and quadrupole form factors, respectively,
calculated by Piarulli and collaborators~\cite{Piarulli12} (magenta hatched bands) and 
Phillips~\cite{Phillips03,Phillips07} (purple bands). Both calculations are based on chiral 
NN potentials. In particular, Piarulli {\it et al.} use wave functions from the chiral
NN potential at N3LO~\cite{Entem03}, while Phillips those from the NN interaction at N2LO~\cite{Epelbaum05}.
The thickness of the bands represents the sensitivity of the results to different 
cutoffs utilized to regularize the divergent behavior at high momenta of the chiral operators'
matrix elements~\cite{Piarulli12,Phillips03,Phillips07}. The two calculations very nicely
agree with the experimental data for low-value of momentum transferred ($q\simeq 3$ fm$^{-1}$) 
and exhibit a  similar (small) cutoff dependence. In the case of the charge form factor,
as $q$ increases,  both theoretical calculations exhibit a more pronounced cutoff dependence,
and differ between each other, an indication that this observable is sensitive to the 
nuclear wave functions utilized in the calculations. In the case of the quadrupole
form factor, the agreement with the experimental data is seen up to $q\simeq6$ fm$^{-1}$,
well beyond the expected regime of validity of the $\chi$EFT framework. In the right panel of
Fig.~\ref{fig-3}, we compare the results for the deuteron magnetic form factor
obtained by Piarulli 
{\it et al.}~\cite{Piarulli12} (hatched magenta band) based on chiral N3LO potential~\cite{Entem03}
and chiral e.m. currents at N3LO~\cite{Pastore09}, with
the fully consistent $\chi$EFT calculations by K\"{o}lling {\it et al.}~\cite{Kolling12} 
(solid purple band) based on the chiral NN potential at N2LO~\cite{Epelbaum05} 
and chiral e.m.~currents at N3LO~\cite{Kolling09,Kolling11}. The theoretical results
are in very good agreement with each other and with the experimental data for values 
of momentum transferred $q\simeq 3$ fm$^{-1}$,  and present a comparable cutoff dependence.

\begin{figure}
\centering
\includegraphics[width=6cm,clip]{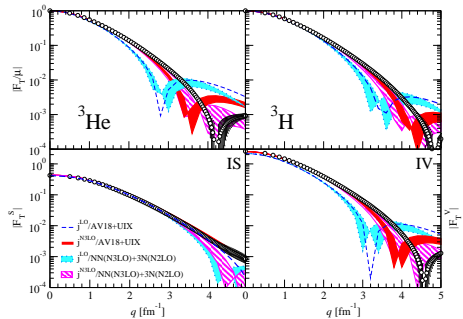}
\caption{The $^3$He and $^3$H  magnetic form factors,
and their isoscalar  and isovector  combinations from Ref.~\cite{Piarulli12}.}
\label{fig-4}       
\vspace*{-0.2in}
\end{figure}

Form factor calculations of $A=3$ nuclei have been reported in Ref.~\cite{Piarulli12}. Here,
we show results for the trinucleon magnetic form factors obtained utilizing the chiral e.m.
currents at N3LO of Refs.~\cite{Pastore08,Pastore09} and two sets of nuclear Hamiltonians, namely
the AV18~\cite{AV18} NN plus UIX~\cite{Pudliner95} 3N potentials, and the N3LO~\cite{Entem03} NN and 
N2LO~\cite{Navratil07b} 3N potentials. Calculations in IA are given in light blue 
(based on chiral interactions) and blue (based on conventional interactions), while
full calculations that include the complete e.m. current up to N3LO are 
given in magenta (based on chiral interactions) and red (conventional interactions).
In the figure, the top panels show the  $^3$He and $^3$H magnetic
form factors, while the bottom ones show their isoscalar ($F_T^S$) and isovector ($F_T^V$) 
combinations~\cite{Piarulli12}. As it is well know from studies based on the conventional 
approach~(see Ref.~\cite{Carlson98}), two-body e.m.~currents are crucial to improve the 
agreement between the observed positions of the zeros and the predicted ones at LO (or IA). 
Despite the excellent agreement between theory and experiment for $q\leq2$ fm$^{-1}$, 
the theory underpredicts the data at higher momentum transfers, while the
zeros are found at lower values of $q$ than observed. The theoretical description of 
the first diffraction region is still incomplete.

\section{Magnetic moments and electromagnetic transitions in $A\le10$ nuclei}
\label{sec-2}

\begin{figure}
\begin{minipage}{0.40\textheight}
\includegraphics[width=5cm,angle=270,keepaspectratio=true]{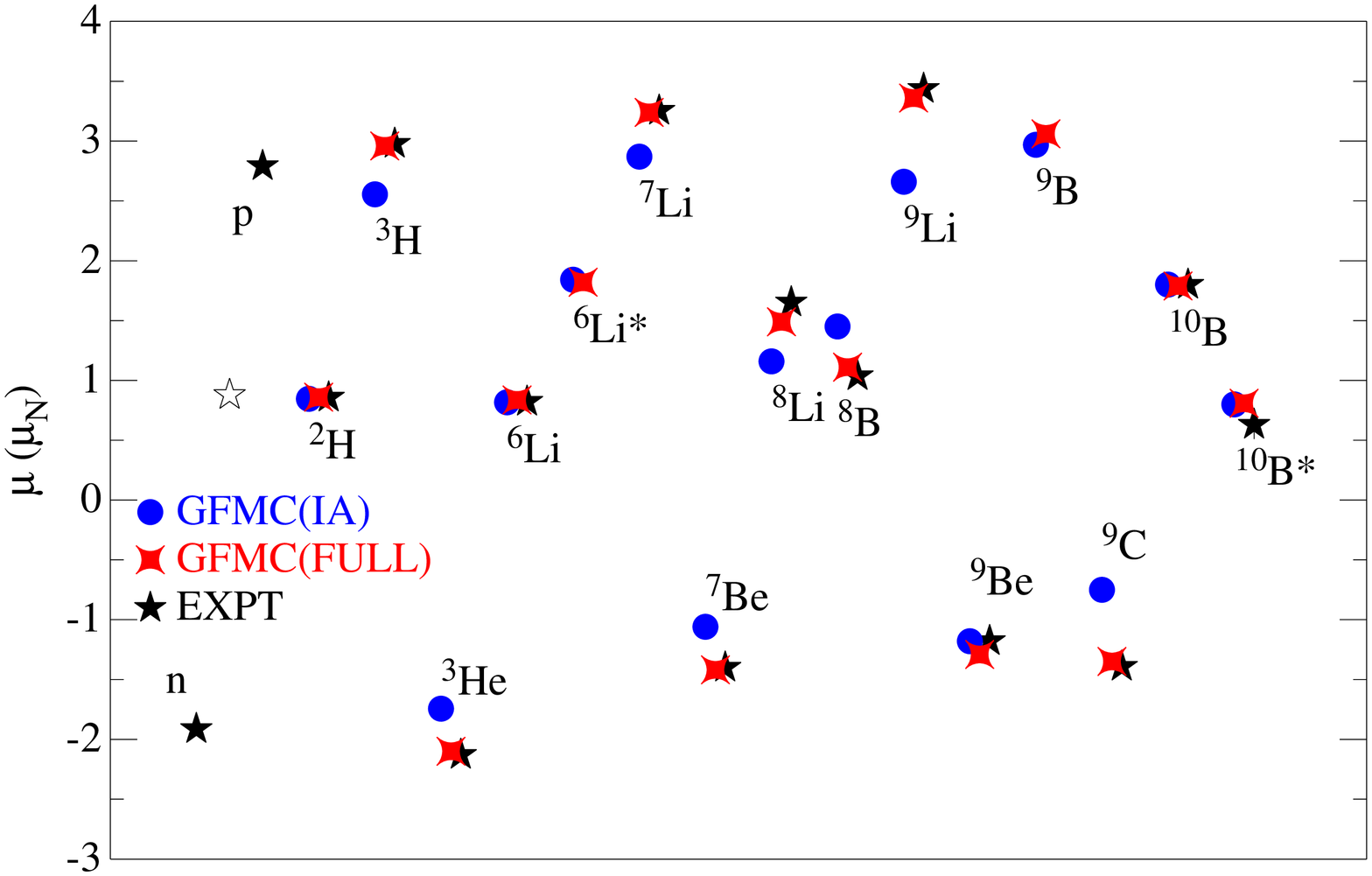}
\end{minipage}
\begin{minipage}{0.40\textheight}
 \hspace*{1cm}
\includegraphics[width=4.5cm]{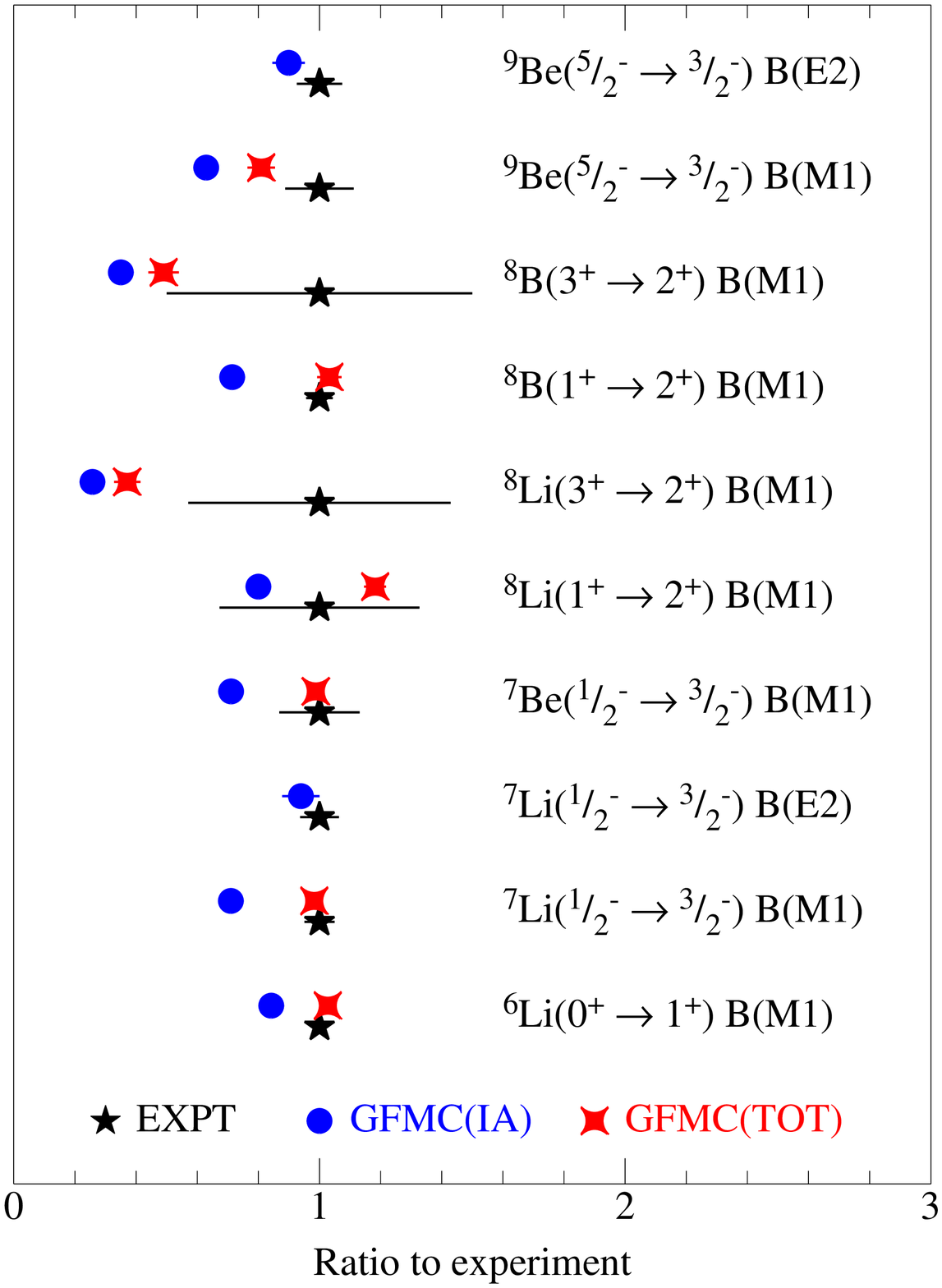}

\end{minipage}
\caption{  {\bf Left}: Magnetic moments in nuclear magnetons for $A\leq10$ nuclei from Ref.~\cite{Pastore12}.
           Black stars indicate the experimental values,
           while blue dots (red diamonds) represent GFMC calculations which include the LO or IA
           (N3LO) e.m.~currents from $\chi$EFT.  
           {\bf Right}: Transition widths from Ref.~\cite{Pastore12} normalized to the experimental values
           for $A=6$--$9$ nuclei. The notation is as in left panel.
           }
\label{fig-5}
\vspace*{-0.2in}
\end{figure}

Moving on to larger nuclear systems, we find a number of Greens's Function Monte Carlo calculations~\cite{Carlson:2014vla}
based on the AV18~\cite{AV18} plus IL7~\cite{IL7} nuclear Hamiltonian that use the chiral 
e.m. currents up to N3LO from Refs.~\cite{Pastore08,Pastore09,Piarulli12}. Magnetic moments of light nuclei~\cite{Pastore12} are summarized in the left panel 
of Fig.~\ref{fig-5}, where IA results are given by blue dots, while calculations that include
the full chiral e.m. current operator are indicated by red diamonds to be compared with
the experimental data represented by black stars. First, we note that corrections from 
two-body currents are found to be small where the IA picture is satisfactory (see, {\it e.g.},
$^6$Li, $^9$Be, $^{10}$B), and large where the IA picture is incomplete (see, {\it e.g.},
$^7$Li, $^7$Be, $^9$C, $^9$Li). Corrections from two-body components can be as large as
40\%, as seen in the case of $^9$C, and are crucial to reach (or improve) the agreement
with the experimental data. It is also interesting to note that  two-body effects, while being
significant for the $^9$C and $^9$Li magnetic moments, are found to be 
negligible for those of $^9$Be and $^9$B. This behavior can be explained  considering
the dominant spatial symmetries of the nuclear wave functions for these $A=9$ systems.
For example, the dominant spatial symmetry of $^9$Be ($^9$B) corresponds to an $[\alpha, \alpha, n (p)]$
structure~\cite{Wiringa06}. Therefore, the unpaired nucleon outside the $\alpha$ clusters
does not interact with other nucleons. As a consequence, two-body currents, that describe the
coupling of external e.m. probes with pair of interacting nucleons, produce a negligible correction.
On the other hand, the dominant spatial symmetry of $^9$C ($^9$Li) corresponds to
an [$\alpha$, $^3$He ($^3$H), $pp$ ($nn$)] structure, and two-body correlations  
contribute in both the trinucleon clusters and in between the trinucleon 
clusters and the valence $pp$ ($nn$) pair, resulting in a large two-body current contribution.

GFMC calculations of selected E2 and M1 transitions in low-lying nuclear states~\cite{Pastore12} are summarized
in the right panel of Fig.~\ref{fig-5}. Predictions in IA are represented by blue dots, while those obtained 
with the full chiral e.m.~current operator are represented by red diamonds. Calculations for E2 transitions
implicitly include the effect of two-body currents via the Siegert theorem, where 
the charge density is used in IA. Also for these observables the effect of two-body e.m. currents
can be large, and for cases in which the experimental errors are relatively small, {\it e.g.}, 
$^7$Li(1/2$^-\rightarrow3/2^-$), $^8$B(1$^+\rightarrow2^+$), 
it is found that their inclusion leads to agreement with the experimental data. This scheme
has been most recently utilized to study e.m. (both E2 and M1) transitions occurring in $^8$Be~\cite{Datar13,Pastore14}.
It is found that the agreement between the calculated and the experimental M1 widths is
not satisfactory. Nevertheless, chiral two-body e.m. currents provide correction 
at the 20\%-30\% level, which, in most but one case, improves the IA values. It
is possible that the systematic underprediction of these observables is due to
a poor knowledge of the small components entering the calculated GFMC nuclear wave
functions~\cite{Pastore14}.

\section{Summary and outlook}
\label{conclusions}

In this talk, I presented an overview on the present status of {\it ab initio} calculations
of e.m. observables, including e.m. moments and form factors, as well as e.m. transitions
in light nuclei. The emphasis was on calculations that account for many-body effects in both 
nuclear Hamiltonians utilized to generate the wave functions, and e.m. current operators.
I focused on results that account for two-body operators that have been derived from
a $\chi$EFT formulation with pions and nucleons,  including up to corrections of two-pion range.
The {\it ab initio} prescription is extremely successful in explaining the experimental data,
provided that many-body effects in both the e.m. currents and nuclear Hamiltonians are
accounted for. $\chi$EFT based calculations of $A=2$ and $3$ nuclei e.m. form 
factors~\cite{Piarulli12} nicely agree with the experimental  
data in the low-energy regime of applicability of $\chi$EFTs, with two-body corrections
playing an important role in improving the agreement between the calculated and the experimental values
of the trinucleon magnetic form factors. E.m. two-body current operators provide a 40\% correction
in the calculated magnetic moment of $^9$C~\cite{Pastore12}, and corrections at the 20\%-30\% level 
in M1 transitions occurring in $^8$Be~\cite{Pastore14}.  There are many interesting e.m. observables
that can be accessed within this formalism. For example, few (or no) {\it ab initio} calculations
of e.m. (charge and magnetic) form factors in $A>4$ nuclei currently exist~\cite{review2014}, and 
it would be interesting to perform them to have a deeper insight on nuclear e.m. structures.
A complete microscopic profile of nuclei includes also studies of e.m. reactions
such as radiative captures and photonuclear reactions. From the theoretical point of view,
 {\it i}) the construction of chiral potentials compatible with Quantum Monte Carlo 
 computational calculations~\cite{Lynn:2014zia}, opens up the possibility of performing
 consistent Quantum Monte Carlo calculations that use chiral potentials and chiral currents;
 {\it ii}) the construction of the chiral NN potential with the explicit inclusion
 of $\Delta$-excitation~\cite{Piarulli:2014bda}, allows study of the effects of $\Delta$-isobars
 in chiral two-body e.m. current operators~\cite{Pastore08}. Weak processes are also 
 being vigorously studied  within the $\chi$EFT formulation. Among these studies are
 the derivation of the axial two-body current operator up to one-loop~\cite{Baroni}, as well as
 the construction of two-body operators entering pion-production reactions
 induced by neutrino scattering off nuclei (see contributions by F.~Myhrer in this proceedings).

\begin{acknowledgement}
This work was supported by the National Science Foundation, grant No. PHY-1068305. I am grateful and owe sincere thanks 
to R.~Schiavilla, R.B.~Wiringa, Steven Pieper, L.~Girlanda, M.~Piarulli, M.~Viviani, L.E.~Marcucci, A.~Kievsky, and S.~Bacca. 
\end{acknowledgement}

%
%
%
\bibliography{bibliography_pastores}

\end{document}